\documentstyle[prl,aps,epsf]{revtex}
\begin{document}
\draft
\twocolumn[
\begin{center}
{\large\bf Non-perturbative approach to correlations in
two-dimensional vortex liquids}\\
\vspace{3ex}
Joonhyun Yeo and M. A. Moore \\
{\em Department of Physics, University of Manchester,
Manchester, M13 9PL, United Kingdom.}\\
(\today )
\end{center}
\widetext
\begin{abstract}
\leftskip 54.8pt
\rightskip 54.8pt
We calculate the renormalized
quartic vertex function of the Ginzburg-Landau model for
a superconducting film in a magnetic field by summing an
infinite subset of diagrams, the so-called
parquet graphs. Using this non-perturbative
solution, we obtain the
structure factor of the two-dimensional vortex liquid.
We find growing
crystalline order in the system as the temperature is lowered.
Our results suggest that the length scale characterizing the crystalline order
diverges only in the zero-temperature limit, which indicates
the absence of a finite-temperature phase transition to the vortex lattice
phase.
\end{abstract}
\leftskip 54.8pt
\rightskip 54.8pt

\narrowtext
]
%%%%%%%%%%
The effect of thermal fluctuations in type II superconductors
in the presence of
a magnetic field has been a focus of theoretical and
experimental interest since the discovery of high $T_c$
superconductors. Mean field theory predicts
the existence of the mixed phase where the magnetic
field penetrates the sample in the form of flux lines
which form a triangular array known as the
Abrikosov vortex lattice \cite{abrikosov}.
Thermal fluctuations are believed to be
responsible for the melting of the Abrikosov vortex lattice
into a vortex liquid state.
The melting of the two-dimensional vortex lattice has been studied
in association with the dislocation unbinding mechanism leading to
a continuous transition \cite{dh}.
However, the question whether the two-dimensional
vortex liquid undergoes a phase transition at all into
the low-temperature ordered state is still controversial.
The high temperature perturbation expansions of the Ginzburg-Landau (GL)
model within the lowest Landau level (LLL) approximation have
been analyzed for
evidence of the transition to the ordered phase at low temperatures
\cite{rt}. Some numerical simulations \cite{sim}
of the same model revealed a weak first-order transition between vortex
liquid-solid phases. On the other hand, the Monte-Carlo
(MC) simulation performed
in Ref.~\onlinecite{om} on a spherical geometry showed no sign of a
finite temperature transition.

In this Letter, we present a non-perturbative analysis of the GL model
for a superconducting film in a magnetic field within the LLL approximation.
We make a resummation of an infinite number
of diagrams summing
all the parquet graphs
\cite{parquet}. We find a clear indication
of growing crystalline order in the two-dimensional
vortex liquid as the temperature is lowered.
However, within our approximation, we find no evidence for
a finite-temperature phase transition.
In fact, our results are consistent with a zero-temperature
scaling argument \cite{om} according to which the length scale
characterizing the spatial order in the vortex liquid diverges
only in the zero temperature limit.

A different kind of non-perturbative scheme for the same model has
been employed by Te\u{s}anovi\'{c}
and co-workers \cite{tetal} to calculate
various thermodynamic quantities.
This method, however, contains an adjustable parameter which is
the generalized Abrikosov ratio, $\beta_A$.
In the present nonperturbative analysis, where no adjustable parameters are
used, all the physical quantities including $\beta_A$ are
calculated as functions of temperature in a straightforward manner.

The starting point of our analysis is
the Ginzburg-Landau free energy for a superconducting film in a
perpendicular magnetic field ${\bf B}={\bf\nabla}\times{\bf A}$
given by
\[
F[\Psi ]=\int d^2{\bf r}\bigg(\alpha |\Psi |^2 +\frac{\beta}{2}
|\Psi |^4+\frac{1}{2m}|(-i\hbar{\bf\nabla} -e^* {\bf A})\Psi |^2\bigg) ,
\]
where $\alpha,\beta$ and $m$ are phenomenological parameters.
For a thin film superconductor, fluctuations in the vector potential
${\bf A}$ can be ignored. We restrict the order parameter $\Psi$ to the
space spanned by the LLL wavefunctions. This approximation
is valid near the upper critical field $H_{c_2}$.
In the symmetric gauge, where ${\bf A}=\case{B}{2}(-y,x)$, the LLL
is fully described by an arbitrary holomorphic function of the
variable $z=x+iy$ multiplied by an exponential factor:
$
\Psi (x,y)=\exp (-\frac{\mu^2}{4}z^* z)\phi(z)$,
where $\mu\equiv \sqrt{e^* B/\hbar }$ is the inverse
magnetic length.
In the LLL approximation, the
GL free energy becomes
\begin{equation}
F[\phi ]=\int dz^* dz\{\alpha_H e^{-\frac{\mu^2}{2}|z|^2}
|\phi (z)|^2+\frac{\beta}{2}e^{-\mu^2|z|^2}|\phi (z)|^4\},
\label{fphi}
\end{equation}
where $\int dz^*dz \equiv\int d^2{\bf r}$
and
$\alpha_H\equiv\alpha+\hbar e^* B/2m=0$ is the point where
the mean field instability occurs. The temperature will
be represented by a dimensionless parameter,
$\alpha_T\equiv\alpha_H\sqrt{2\pi/\beta\mu^2}$.
In the high temperature perturbation theory, one uses
$x\equiv \mu^2\beta/4\pi\alpha^2_R$ as a dimensionless
expansion parameter, where
$\alpha_R$ denotes
the renormalized $\alpha_H$ ({\it i.e.} the
denominator of the $\langle\Psi^*\Psi\rangle$ propagator).
The zero-temperature limit corresponds to
$\alpha_T\rightarrow -\infty$ or $x\rightarrow\infty$.
The parameter $\alpha_R$ is usually estimated via the Hartree
approximation, where $\alpha_R =\alpha_H (
1-4x)^{-1}$ \cite{rt}. Using our non-perturbative method,
we will be able to obtain a much improved estimate for $x$ and
$\alpha_R$.

In the renormalization group analysis of the
generalization of (\ref{fphi}) in $6-\epsilon$ dimensions,
Br\'ezin {\it et al.} \cite{bnt} indicated that
the GL hamiltonian in (\ref{fphi}) is not closed under renormalization.
The renormalization drives the quartic vertex of (\ref{fphi})
to the most general gauge invariant form,
$
\int\prod_{i=1,2}dz^*_i dz_i e^{-\frac{\mu^2}{2}(|z_1|^2
+|z_2|^2)}|\phi(z_1)|^2 g(|z_1 -z_2|)|\phi (z_2)|^2,
$
for an arbitrary function $g$.
In this Letter, we use the Fourier transform representation \cite{mn}
of the quartic vertex instead of the one used in Ref.~\onlinecite{bnt}:
$\widetilde{g}(k)=\int d^2{\bf r}\; g (r)\exp(i{\bf k}\cdot{\bf r})$.
It is convenient for later
use to introduce the scaled function $f(k)\equiv (2/\beta)\exp
(-k^2/2\mu^2  )\widetilde{g}(k)$. Note that the bare interaction
corresponds to $f_B (k)=\exp(-k^2/2\mu^2)$.

In the present work, we obtain the renormalized quartic vertex function,
$f_R(k)$
of the GL hamiltonian by summing all the parquet graphs
\cite{parquet}.
We then calculate the structure factor of the two-dimensional
vortex liquid from this non-perturbative
solution. The structure factor denoted by $\Delta(q)$ is proportional to the
Fourier transform of the
spatial correlation function of the superfluid density, $
\widetilde{\chi}(q)=\int d^2{\bf R}\;e^{i{\bf q}\cdot {\bf R}}\chi
({\bf r},{\bf r}+{\bf R}),$ where
$\chi({\bf r},{\bf r}^\prime)\equiv\langle |\Psi ({\bf r})|^2
|\Psi ({\bf r}^\prime)|^2\rangle -\langle |\Psi ({\bf r})|^2\rangle
\langle |\Psi ({\bf r}^\prime)|^2\rangle $:
\begin{equation}
\widetilde{\chi}(q)=\frac{\mu^2}{2\pi\alpha^2_R}e^{-q^2/
(2\mu^2)}\Delta (q).
\label{chiq}
\end{equation}
The structure factor of the two-dimensional vortex liquid has been studied
using the
numerical simulations \cite{sim}, the density-functional theory
of vortex liquid freezing \cite{dft} and the high-temperature
perturbation theory \cite{hmm}. In terms of $f_R(k)$,
$\Delta(q)$ is given by
\begin{equation}
\Delta (Q)=1-2x\{ f_R (Q)+\int^\infty_0 dK K J_0 (QK)f_R (K)\} ,
\label{deltaq}
\end{equation}
where $J_0$ is the Bessel function and
${\bf Q}={\bf q}/\mu$ and ${\bf K}={\bf k}/\mu$ are dimensionless
wave vectors. Eq.~(\ref{deltaq})
is the central quantity of our investigation.
In (\ref{deltaq}), the first term
comes from contracting two pairs of external lines
in the disconnected four-point
Feynman diagram. The remaining terms represent the two independent ways
of contracting the connected four-point diagrams.

Using $f_R (K)$, one can also express $\alpha_R$ in terms of $\alpha_H$.
The relation between the renormalized parameter $\alpha_R$
(or $x$) and $\alpha_H$ (or
$\alpha_T)$ gives the
specific heat of the system.
In the present analysis, this relation comes from
the Dyson-Schwinger equation associated with the hamiltonian
(\ref{fphi}), which expresses the renormalized propagator
in terms of the bare propagator and the renormalized
four-point function.
It follows that
\[
\alpha_T=\frac{1}{\sqrt{2x}}\Big[1-4x\{ 1-2x\int_0^\infty
dK\; K\; f_R (K)e^{-K^2/2}\} \Big] .
\]
The Hartree approximation therefore corresponds to neglecting the term
that depends on $f_R(K)$.
The generalized
Abrikosov ratio is defined by
$
\beta_A(x)\equiv\overline{\langle |\Psi ({\bf r})
|^4\rangle}/[\overline{\langle |\Psi ({\bf r})|^2
\rangle} ]^2$, where the bar denotes the spatial average.
In terms of $f_R(K)$, we have
\begin{equation}
\beta_A(x)=
2\Big[ 1-2x\int_0^\infty dK K f_R (K) e^{-K^2/2}\Big] .
\label{betaa}
\end{equation}
It is interesting to see
if $\beta_A(x)$ really converges to
$1.16$ as $x\rightarrow\infty$ or $\alpha_T\rightarrow -\infty$.

In order to calculate $f_R(K)$, we
consider the Feynman diagrams contributing to the connected
four-point function, $
W (z^*_1,z^*_2,z_3,z_4)\equiv\langle\phi^* (z^*_1)
\phi^* (z^*_2)\phi(z_3 )\phi(z_4)\rangle_c$.
We then make the parquet graph analysis on $W$ and obtain the
corresponding relations for $f_R(K)$ to which $W$ is directly
related within the LLL approximation \cite{bnt,mn}.
The four-point vertex function depends
on three variables characterizing the three independent channels.
Since $f_R(K)$ is defined through
the faithful representation of diagrams,
we express all the diagrams involved in the parquet decomposition for $W$
in the faithful representation.

Since an arbitrary diagram contributing $W$ is either
reducible or irreducible for a given channel $i=1,2$ or $3$,
one can write in terms of $f_R(K)$ that
$f_R (K)=I_i(K)+\Gamma_i(K)$, where $I_i$ and
$\Gamma_i$ are the irreducible and reducible parts for channel $i$,
respectively. The irreducible part $I_i(K)$ consists of the bare
vertex function $f_B(K)$, the totally irreducible part $R(K)$, which
is irreducible in all channels, and those which are reducible
in the other two channels. Thus, we have
\begin{equation}
I_i(K)=f_B(K)+R(K)+\sum_{j\neq i}\Gamma_j (K).  \label{ii}
\end{equation}
Therefore,
\begin{equation}
f_R(K)=f_B(K)+R(K)+\sum_{i=1}^{3}\Gamma_i(K). \label{g123}
\end{equation}
\begin{figure}
\centerline{
\epsfxsize=8cm \leavevmode \epsfbox{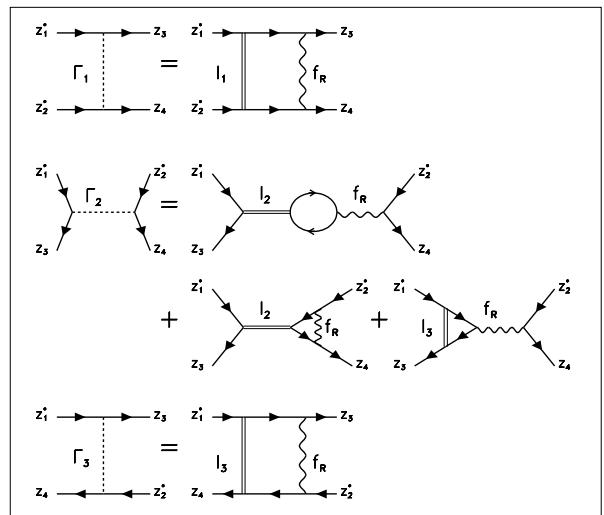}}
\caption{Dia\-gramm\-atic rep\-resen\-ta\-tion of the Bethe-Sal\-peter
equa\-tion
for the re\-duci\-ble parts $\Gamma_i (K)$.}
\label{fig:bs}
\end{figure}
\noindent
The reducible part $\Gamma_i (K)$ can be represented in terms of
the Bethe-Salpeter equation as described graphically in Fig.~\ref{fig:bs}.
Essentially, they express the reducible part
in terms of a series of ladder graphs
composed of appropriate irreducible parts.
The detailed derivation of these relations will be presented in a separate
publication \cite{ym}.
By explicitly evaluating
each diagram in Fig.~\ref{fig:bs}, we obtain
\begin{mathletters}
\label{g1g2g3}
\begin{eqnarray}
\Gamma_1 (K)&=&-2x\int\frac{d^2{\bf P}}{2\pi}
I_1(|{\bf K}-{\bf P}|) \\
&&~~~~~~~~~~~~~~~~~\times f_R(P) \cos ({\bf K}\times{\bf P}) \nonumber \\
\Gamma_2 (K)&=&-2x\{ I_2(K)f_R(K) \\
&&~~~~~+I_2 (K)\int
\frac{d^2{\bf P}}{2\pi} f_R (P)\cos ({\bf K}\times{\bf P}) \nonumber \\
&&~~~~~+
f_R(K)\int\frac{d^2{\bf P}}{2\pi}\; I_3 (P)\cos({\bf K}\times
{\bf P})\} \nonumber \\
\Gamma_3 (K)&=&-2x\int\frac{d^2{\bf P}}{2\pi}\;I_3(
|{\bf K}-{\bf P}|)f_R(P),
\end{eqnarray}
\end{mathletters}
$\!\!$where ${\bf K}\times{\bf P}\equiv K_1 P_2 -K_2 P_1$.
It is important to note that the coupled integral equations,
(\ref{g123}) with (\ref{ii}) and (\ref{g1g2g3})
are {\em exact} relations for $f_R (K)$.
In the present work, we neglect the contribution from
the nonparquet diagrams
represented by $R(K)$ in (\ref{ii}), for which
a systematic analysis is very difficult.
We note that the lowest order diagram contributing to $R(K)$ is of
$O(\beta^4)$.
The subset of
parquet graphs contain an enormous number of diagrams and we
assume that the essential properties of the vortex liquid
can be represented by this subset.
We solve numerically eqs.
(\ref{g123}), (\ref{ii}) and (\ref{g1g2g3}) for $f_R(K)$
with the condition $R(K)=0$
and evaluate $\Delta (K)$ using (\ref{deltaq}).
For given set
of functions,
$\{ I_i (K)\}$, Eq.~(\ref{g123})
is a linear equation for $f_R(K)$
(a Fredholm equation of the second kind) which can be
solved by a numerical inversion of a matrix
\cite{numericalrecipe}.
The solution is then found by iteration with the improved irreducible
parts calculated from (\ref{ii}) using the solution to the linear equation
just obtained.

\begin{figure}
\centerline{
\epsfxsize=7cm \leavevmode \epsfbox{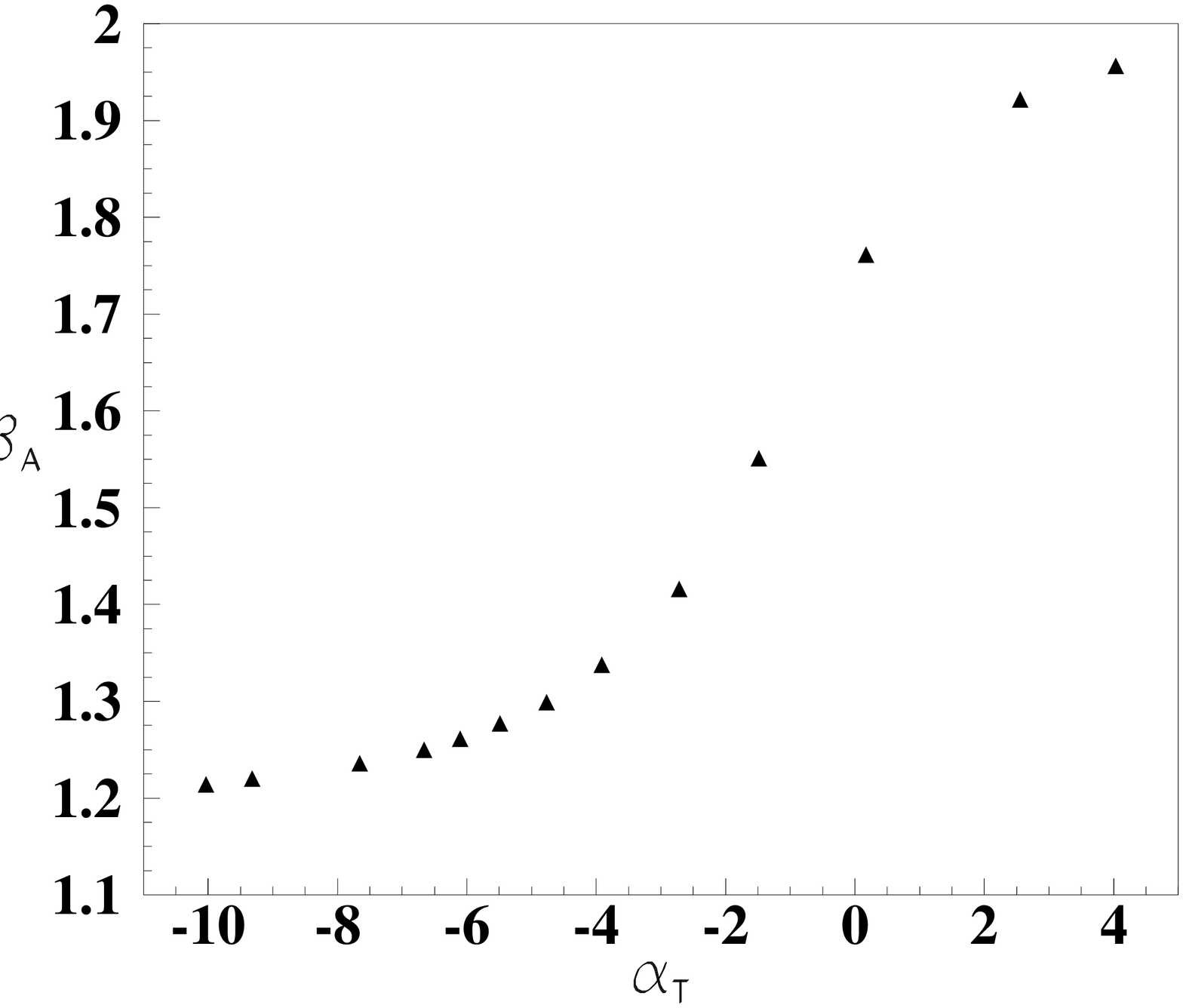}}
\caption{The ge\-nera\-lized Ab\-riko\-sov ratio as a func\-tion
of tem\-pera\-ture.}
\label{fig:abrikosov}
\end{figure}

We obtain $f_R(K)$ and $\Delta(K)$ for various values of $x$.
Before presenting our results for $\Delta (K)$, we first discuss
how the perturbative expansion parameter $x$ depends on
dimensionless temperature $\alpha_T$.
{}From our numerical data, we find a significant improvement over the
Hartree approximation.
We also find that
at low temperatures the
$\sqrt{x}$ vs. $\alpha_T$ curve approaches a
straight line, the slope of which is related to the Abrikosov ratio
in the zero temperature limit.
We can in fact plot directly the Abrikosov
ratio as a function of temperature using (\ref{betaa}). (See
Fig.~\ref{fig:abrikosov}.) The value of
$\beta_A$ decreases with decreasing temperature from its high temperature
limit $\beta_A=2$ representing the
uncorrelated vortex liquid.
As one goes into the low temperature region,
$\beta_A$ decreases very slowly. Although it is hard to determine
the exact asymptotic value at
$\alpha_T=-\infty$ from the present data,
these values are getting close
to $1.16$, the value for a triangular lattice.

\begin{figure}
\centerline{
\epsfxsize=8cm \leavevmode \epsfbox{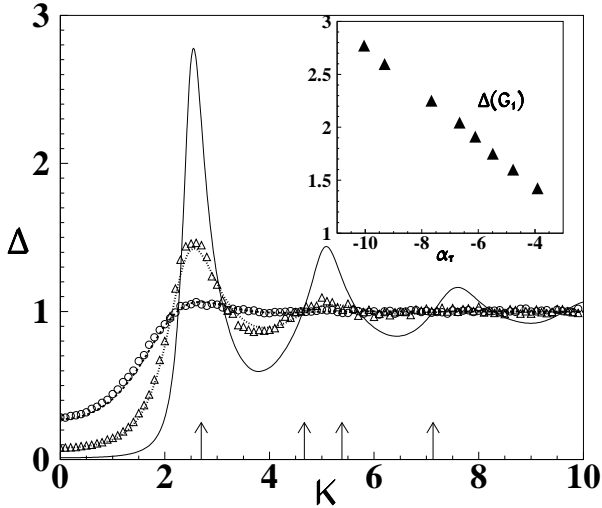}}
\caption{The structure factor $\Delta (K)$ calculated from the present
non-perturbative method for various temperatures;
$\alpha_T= -1.5$ (the broken line),
$-3.9$ (the dotted line) and $-10$ (the solid line). The arrows indicate the
positions of the RLVs of the triangular lattice. The open circles and
the open triangles are those obtained in the MC simulation
for $\alpha_T=-1.5$ and $-4.0$, respectively. The inset shows the height of
the first peak obtained from the value of $\Delta$ at the first RLV, $G_1$
as a function of temperature.}
\label{fig:delta}
\end{figure}

Fig.~\ref{fig:delta} shows the structure factor $\Delta(K)$
for various temperatures. We find that, as $\alpha_T\lesssim -2$,
the structure factor
starts to develop a well-defined peak
near the first reciprocal lattice vector (RLV) of a triangular
lattice, $G_1/\mu\simeq 2.69$.
As the temperature is lowered further, more structures are
revealed as the peaks around the next RLVs begin to appear.
Since the second and third RLVs of the
triangular lattice are closely spaced, our solution could not resolve
these peaks up to the minimum temperature we have considered.
However, one can clearly see that the height of the first peak grows
while the width narrows down as the temperature is lowered,
which indicates the growing crystalline order in the
two-dimensional vortex liquid system.

In Fig.~\ref{fig:delta}, our nonperturbative results for $\Delta(K)$
are compared with the recent MC simulation data \cite{mattd}.
Up to low
temperatures, they are in good agreement with the MC results.
But we find that, as one goes further into the low temperature regime,
the peak height in the MC results grows faster than in the present
calculation.
It is to be expected that the diagrams omitted from the parquet
approximation will produce quantitative errors at low temperatures,
whereas at high temperatures their effects are negligible.
In Ref.~\onlinecite{hmm}, the structure factor of the
two-dimensional vortex liquid has been evaluated up to 12th order of
perturbation theory. The extrapolation to low temperatures
via Pad\'e approximants shows a peak
starting to appear near the first RLV
of the triangular lattice, but the method becomes
unreliable at low temperatures
and at small wave vectors.

In the completely ordered state where vortices form a lattice,
the structure factor would exhibit delta-function like
divergences at the reciprocal lattice vectors (RLV).
In fact, if we use the mean field solution for $\Psi({\bf r})$
\cite{abrikosov} in the definition of the structure factor, we get
$\Delta ({\bf k})=(2\pi)^2\sum_{{\bf G}\neq 0}\delta^{(2)}
({\bf k}-{\bf G})$, where the sum is over the RLVs
of the
triangular lattice except ${\bf G}=0$. The ${\bf G}=0$ term corresponds to the
disconnected piece substracted in the
\noindent
definition of $\chi$.
Assuming that $\Delta(K)$ obtained
from our calculation approach
this low temperature limit, we
investigate the temperature dependence
of the height of the growing peaks in the structure factor.
{}From our numerical solution, we find that in the low temperature regime
the height of the first peak grows as $|\alpha_T|$. (See the inset in
Fig.~\ref{fig:delta}.) This result can be understood in terms of
the zero-temperature scaling argument \cite{om} mentioned before.
This argument is based on the observation that
the destruction of off-diagonal long range order below four dimensions
\cite{moore} by thermal
fluctuations
might be accompanied by the destruction of the positional order
characterized by a length scale $\xi$. In this argument, $\xi$
diverges in the zero-temperature limit as $|\alpha_T|$. If we can describe
$\Delta (k)$ near a peak $G$ by the function
$\xi h(\xi (k-G))$,
then the integral of
$\Delta(k)$ over ${\bf k}$ around $G$ is finite as $\xi\rightarrow\infty$
\cite{ym},
which is required by the asymptotic form of $\Delta (k)$. Therefore, the
peak height at $k=G$
grows as $\xi\sim |\alpha_T|$ as $\alpha_T\rightarrow -\infty$.

We note that our non-perturbative approximation
captures this asymptotic behavior suggested by the zero temperature scaling.
{}From (\ref{deltaq}), we deduce that in the zero temperature limit,
$f_R({\bf k})\sim (2x)^{-1}\{ 1-\pi\mu^2\sum_{{\bf G}}\delta^{(2)}
({\bf k}-{\bf G})\}$. If we put this expression into eqs.~(\ref{ii}),
(\ref{g123}) and (\ref{g1g2g3}) and focus on the region
${\bf k}\sim {\bf G}\neq 0$, we find $(4\pi/x)\delta^{(2)}
({\bf k}-{\bf G})=$ a smooth function of
${\bf k}$ around ${\bf G}$ independent of $x$.
Upon integrating over the region with the
peak width $\xi^{-1}$, we find $\xi^2\sim x\sim\alpha^2_T$ as
$\alpha_T\rightarrow -\infty$. This argument indicates that our data
for $\Delta(K)$ already exhibit the asymptotic behavior suggested by the
zero-temperature scaling and therefore within our approximation the
phase transition to the vortex lattice state at finite temperature is
highly unlikely.

To summarize, we have found an analytic approach to the vortex liquid
regime which is sophisticated enough to predict, {\it ab initio}, the
growth of crystalline order as the temperature is lowered; which agrees
qualitatively with MC estimates of the structure factor at moderate to high
temperatures; while in the low-temperature regime it is consistent with
zero-temperature scaling argument, and the absence of a finite
temperature phase transition.

We thank T.Newman for useful
discussions and M.Dodgson for providing his MC simulation data.

\end{document}